# Ideal spectral emissivity design for extreme radiative cooling

Suwan Jeon[1] and Jonghwa Shin[1,a]

[1]*Department of Materials Science and Engineering, Korea Advanced Institute of Science and Technology, Daejeon 34141, Republic of Korea*

**Abstract**

Radiative coolers that can passively cool objects by radiating heat into the outer space have recently received much attention. However, the ultimate limits of their performance as well as their ideal spectral design are still unknown. We present the fundamental lower bound of the temperature of a radiatively cooled object on earth surfaces under general conditions, including non-radiative heat transfer, and the upper bound of the net radiative power density of a radiative cooler as a function of temperature. We derive the ideal spectral emissivities that can realize such bounds and, contrary to common belief, find that the ideal emission window is different from 8 to 13 μm and forms disjointed sets of wavelengths, whose width diminishes at lower temperatures. We show that ideal radiative coolers with perfect thermal insulation against conduction and convection have a steady-state temperature of 243.6 K in summer and 180.5 K in winter, much below previously measured values. We also provide the ideal emission window for a single-band emitter and show that this window should be much narrower than that of previous designs if the objective is to build a radiative freezer that can operate in summer. We provide a general guideline for designing spectral emissivity to achieve the maximum temperature drop or the maximum net radiative power density.

**Manuscript**

The atmosphere acts as a selective window to electromagnetic waves, transmitting most of the visible sunlight and blocking harmful ultraviolet and X-rays[1]. Recently, the transparency window in the long-wavelength infrared, which ranges from 8 to 13 μm, has received great attention because it allows for below-ambient cooling without energy consumption[2-15]. The principle of this spontaneous cooling is based on thermal radiation in the transparency window being able to transfer heat between an object on the earth surface and the cold outer space without being blocked by the atmosphere. Even under direct sunlight, below-ambient cooling has been experimentally achieved by enhancing the reflectivity in the solar spectrum and the emissivity in the transparency window using various methods[3,7,9–12,15]. However, it remains unproven whether the 8.0–13.0 μm range is the exact optimal range to realize the maximum temperature drop versus ambient temperature, and the fundamental limit of the achievable cooling temperature has not been theoretically derived yet.

Here, we present the ideal spectral emissivity under general conditions that realizes the ultimate lower bound of the temperature of a radiatively cooled object on earth. Contrary to common belief, we prove that the optimal window is different from 8 to 13 μm and depends strongly on the target temperature, which in turn is limited by the non-radiative heat transfer rate. Particularly, with no non-radiative heat transfer, the cooler should have a needle-like spectral emissivity if the objective is to have as low steady-state temperature as possible, and can induce 56.4 K and 92.5 K drops in summer and winter, respectively, if enough time is given to reach a steady state. More generally, we establish a universal guideline for designing spectral emissivity under given environmental conditions, including ambient temperature and the non-radiative heat transfer coefficient, and present the ultimate lower bound of the temperature as well as the ultimate upper bound of the net radiative power density under those conditions.

___________________

[a] E-mail: qubit@kaist.ac.kr

Before we derive the ideal spectral emissivity, we provide our model and its assumptions on how a radiative cooler interacts with its surroundings. A radiative cooler that covers a target object is horizontally placed on the ground at sea level with no nearby obstacles so that the view angle of the sky is $2\pi$ steradian. The cooler and the object are in thermal equilibrium and thus have the same temperature ($T$). The cooler undergoes radiative heat exchange with the sun, the outer space, and the atmosphere. It also undergoes non-radiative, i.e., conductive and convective, heat exchange with the ambient air and the ground (Fig. 1(a)). The radiation through the atmosphere is subject to various conditions, such as season[16], humidity[17], and temperature variations along the altitude[18]. We denote these environmental variables as $\alpha$. The ground is assumed to be at the same temperature as the ambient air ($T_{amb}$). Then, the net cooling power density of the radiative cooler is given by

$$P_{net}(T_{amb}, T, \alpha) = P_{cooler}(T) - P_{sun}(T_{amb}, T, \alpha) - P_{space}(T_{amb}, T, \alpha) - P_{atm}(T_{amb}, T, \alpha) - P_{non-rad}(T_{amb}, T), \quad (1)$$

where $P_{cooler}(T)$ is the radiant exitance of the cooler, $P_{sun}(T_{amb}, T, \alpha)$, $P_{space}(T_{amb}, T, \alpha)$, and $P_{atm}(T_{amb}, T, \alpha)$ are the absorbed irradiance on the cooler from the sun, the outer space, and the atmosphere, respectively, and $P_{non-rad}(T_{amb}, T)$ is the absorbed non-radiative power density from the surroundings; these all have a unit of W/m². $P_{space}(T_{amb}, T, \alpha)$ can be usually ignored because the cosmic background is much cooler than $T_{amb}$ or $T$. The rest of the terms in Eq. (1) are expressed as

$$P_{cooler}(T) = \int d\Omega \cos\theta \int d\lambda \, \tilde{I}_{BB}(\lambda, T) \epsilon_c(\lambda, \Omega, T), \quad (2a)$$

$$P_{sun}(T_{amb}, T, \alpha) = \int d\lambda \, I_{sun}(\lambda, \Omega_{sun}, T_{amb}, \alpha) \epsilon_c(\lambda, \Omega_{sun}, T), \quad (2b)$$

$$P_{atm}(T_{amb}, T, \alpha) = \int d\Omega \cos\theta \int d\lambda \, \tilde{I}_{atm}(\lambda, \Omega, T_{amb}, \alpha) \epsilon_c(\lambda, \Omega, T), \quad (2c)$$

$$P_{non-rad}(T_{amb}, T) = h_c(T_{amb} - T), \quad (2d)$$

where $\int d\Omega = \int_0^{2\pi} d\phi \int_0^{\pi/2} d\theta \sin\theta$ is the hemispherical integration and $\tilde{I}_{BB}(\lambda, T) = \frac{2hc^2}{\lambda^5} \frac{1}{\exp(hc/\lambda k_B T) - 1}$ is the spectral radiance of an ideal blackbody following Plank's law ($h$, $c$, $k_B$, and $\lambda$ are the Plank constant, the velocity of light in vacuum, the Boltzmann constant, and wavelength, respectively). $\epsilon_c(\lambda, \Omega, T)$ represents the spectral and directional emissivity of the cooler. $I_{sun}(\lambda, \Omega_{sun}, T_{amb}, \alpha)$ is the spectral solar irradiance at a mid-latitude, sea-level location in the northern hemisphere when the sunlight is incident from angle $\Omega_{sun}$. In Eqs. (2b-c), the absorptivity of the cooler is replaced by its emissivity using Kirchhoff's law. In Eq. (2d), the non-radiative absorption is expressed by an effective non-radiative heat transfer coefficient $h_c$, which depends on environmental conditions, such as wind speeds, as well as on how well the cooler and the object are thermally insulated from the environment[19]. $\tilde{I}_{atm}(\lambda, \Omega, T_{amb}, \alpha)$ is the angle-dependent spectral radiance from the atmosphere, obtained by multiplying the black-body radiation spectrum with the atmospheric emissivity $\epsilon_{atm}$. The latter has been approximated as $\epsilon_{atm}(\lambda, \Omega, T_{amb}, \alpha) = 1 - t(\lambda, T_{amb}, \alpha)^{AM(\theta)}$[20], where $t(\lambda, T_{amb}, \alpha)$ is the atmospheric transmittance from the sea level toward the outer space in the zenith direction (data from MODTRAN 6[21]) and AM($\theta$) accounts for larger attenuations at a non-zero zenith angle $\theta$. In doing so, we applied a spherical shell model for AM($\theta$) (Fig. S1 in the supplementary material) instead of the flat-earth model of AM($\theta$) = $1/\cos\theta$[20].

Now, let us derive the ideal spectral emissivity of a radiative cooler at a specific temperature for maximum net cooling power density assuming it is isotropic at all angles. We first rearrange the net cooling power density in Eq. (1) into radiative and non-radiative parts as

$$P_{net}(T_{amb}, T, \alpha) = P_{rad}(T_{amb}, T, \alpha) - P_{non-rad}(T_{amb}, T)$$
$$= \int d\lambda \, [I_{BB}(\lambda, T) - I_{sun}(\lambda, \Omega_{sun}, T_{amb}, \alpha) - I_{atm}(\lambda, T_{amb}, \alpha)] \epsilon_c(\lambda, T) - h_c(T_{amb} - T), \quad (3)$$

where $I_{BB}(\lambda, T) = \int d\Omega \cos\theta \, \tilde{I}_{BB}(\lambda, T)$ and $I_{atm}(\lambda, T_{amb}, \alpha) = \int d\Omega \cos\theta \, \tilde{I}_{atm}(\lambda, \Omega, T_{amb}, \alpha)$ are spectral irradiances and $P_{rad}(T_{amb}, T, \alpha) = P_{cooler}(T) - P_{sun}(T_{amb}, T, \alpha) - P_{atm}(T_{amb}, T, \alpha)$ is the net radiative power density. We note that $\epsilon_c(\lambda, T)$ works as a scaling factor for the net radiative heat exchange at each wavelength. For maximal $P_{net}(T_{amb}, T, \alpha)$ and $P_{rad}(T_{amb}, T, \alpha)$ at a given temperature, $\epsilon_c(\lambda, T)$ can be suitably adjusted between 0 and 1 depending on the sign of the term $I_{rad,BB} = I_{BB}(\lambda, T) - I_{sun}(\lambda, \Omega_{sun}, T_{amb}, \alpha) - I_{atm}(\lambda, T_{amb}, \alpha)$. For wavelengths at which the sign is positive (i.e., net radiative emission occurring), $\epsilon_c(\lambda, T) = 1$ should be chosen to maximize the radiation whereas, for other wavelengths, $\epsilon_c(\lambda, T) = 0$ should be chosen to minimize absorption. This emissivity design rule is summarized as

$$\epsilon_{ideal}(\lambda; \Omega_{sun}, T_{amb}, T, \alpha) = \frac{1}{2}[1 + \text{sgn}(I_{rad,BB})], \tag{4}$$

where $\text{sgn}(x) = \begin{cases} 1 \text{ if } x > 0 \\ -1 \text{ if } x < 0 \end{cases}$ is the sign function, whose value at $x = 0$ is undefined but does not affect the results in this situation.

We emphasize that $\epsilon_{ideal}$ is sensitively determined by the temperature of the cooler. This aspect was not fully investigated in previous studies, and two types of emissivity patterns were usually considered as optimal: a broadband emissivity pattern for above-ambient cooling and a rectangular (1 over 8–13 μm and 0 for other wavelengths) emissivity pattern for below-ambient cooling[6,7,14,22–25]. However, the actual spectral regions contributing to cooling and heating via radiation change dramatically as the temperature drops below the ambient temperature. Thus, emissivity spectra designed for a temperature range around ambient temperature are no longer an optimal solution at lower temperatures.

Another important fact revealed by Eq. (4) is that the temperatures achievable through radiative cooling have a fundamental lower bound. For $P_{rad}(T_{amb}, T, \alpha)$ to be positive, $I_{rad,BB}$ must be positive at least for one wavelength. This imposes the ultimate lower bound, $T_{ideal,min}$, for the temperature of a radiatively cooled object, where $T_{ideal,min}$ is the temperature at which $I_{rad,BB}$ is zero at one (or more) wavelengths and negative at all other wavelengths.

The temperature of the cooler in a steady state can be identified by solving Eq. (3) for $P_{net}(T_{amb}, T, \alpha) = 0$. First, we investigate the most extreme case of $h_c = 0$. Without non-radiative heat transfer, $P_{net}(T_{amb}, T, \alpha)$ is always positive if $T > T_{ideal,min}$. Thus, the ideal radiative cooler cools down until the temperature approaches $T_{ideal,min}$. For example, at Daejeon city (36.35° N in latitude), $T_{ideal,min}$ is 243.6 K and 180.5 K at noon in summer and winter, respectively, as shown in Figs. 1(b–c). This corresponds to 56.4 K and 92.5 K drops below ambient temperature ($T_{amb}$ is assumed to be 300 K and 273 K in summer and winter, respectively). Owing to seasonal variations in atmospheric emissivity, the temperature drop is much larger in winter than in summer. As expected from Eq. (4), the ideal emissivity at these temperatures is a needle-like function centered at 8.96 μm in summer and 11.61 μm in winter; these wavelengths are the only ones at which $I_{rad,BB}$ is non-negative. Considering the lowest measured temperature in previous studies (approximately 243 K under vacuum conditions in winter)[5], our results suggest that temperature can be further lowered by more than 50 K in principle.

In reality, the presence of conduction or convection ($h_c \neq 0$) increases the steady-state temperature of an ideal cooler, $T_{ideal}$, above $T_{ideal,min}$. The steady-state temperature of any radiative cooler with temperature-independent spectral emissivity can be easily found graphically because $P_{rad}$ and $P_{non-rad}$ are monotonically increasing and decreasing functions of $T$, respectively, and they cross each other at the steady-state temperature. For example, $P_{rad}$'s for an 8–13 emitter ($\epsilon_{8-13}$, black dashed line) and a broadband emitter ($\epsilon_{Full}$, black dotted line) are plotted in Fig. 2(a), under summer atmospheric conditions assuming solar irradiance corresponding to AM1.5 and an average solar zenith angle of 48.2°. The 8–13 emitter and the broadband emitter have a unity emissivity for wavelengths between 8 and 13 μm and for all wavelengths longer than 4 μm, respectively. Both hypothetical emitters have exactly zero emissivity for all other wavelengths. The red solid lines in Fig. 2(a) represent $P_{\text{non-rad}}$ for several $h_c$ values. As expected, the steady-state temperature for either emitter rises for higher $h_c$ values.

In general, the net radiative power density of any radiative cooler cannot exceed the values indicated by the black solid line in Fig. 2(a), representing the performance of an ideal radiative cooler at each cooler temperature. In previous studies, the 8–13 emitter was considered as an almost ideal radiative cooler for below-ambient cooling cases. However, it can be seen that the net radiative power density of this cooler (black dashed line in Fig. 2(a)) never touches the black solid line at any temperature. In other words, there is always a better spectral design than unit emissivity from 8 to 13 μm at any target temperature. For example, at 273.15 K, the net radiative power density of the 8–13 μm emitter is 7.31 W/m², whereas that of the ideal cooler optimized for this temperature is more than two times larger at 16.84 W/m². Note that these results are independent of $h_c$. For the particular case of $h_c$ = 0.5 W/(m²K), the lower bound of the steady-state temperature, which is 271.10 K, is below the freezing temperature of water. The ideal cooler, which can reach this bound, has a highly selective spectral emissivity with many disjointed sets of wavelengths over which the emissivity is unity, as shown in Fig. 2(b). Thermal insulation on the level of $h_c$ = 0.5 W/(m²K) can be achieved with a 7 cm-thick polystyrene foam at the back and an infrared-transmitting composite window at the front. Inside the previously accepted 8–13 μm transparency window, it is possible to identify several important wavelength ranges for which the emissivity must be minimized. One such wavelength range is around 9.5 μm, where there are multiple atmospheric absorption resonances due to $O_3$ and other substances[23]. If these wavelengths are included in the emission band as done in previous designs, the steady-state temperature (275.84 K) rises above the freezing point of water.

At lower $h_c$ values, the difference becomes more dramatic. $T_{\text{ideal}}$ and $\epsilon_{\text{ideal}}$ are plotted for a range of $h_c$ values in Figs. 2(c–d). As $h_c$ decreases, the difference between $T_{\text{ideal}}$ and the steady-state temperature of the 8–13 μm emitter ($T_{8-13}$) becomes larger, reaching 24.67 K at $h_c$ = 0, as shown in Fig. 2(c). The width of the wavelength ranges for which the emissivity should be unity diminishes as $h_c$ is reduced and become highly selective, as illustrated in Fig. 2(d). These results imply that, in a properly insulated system, it is possible to achieve a much lower steady-state temperature than $T_{8-13}$ if the spectral emissivity is optimally designed to benefit from the lower $h_c$.

In practice, however, it might be challenging to realize such a highly selective spectrum. Thus, we also consider a simper, single-band emitter with unit emissivity over a single wavelength range from $\lambda_{\text{short}}$ to $\lambda_{\text{long}}$ and optimize $\lambda_{\text{short}}$ and $\lambda_{\text{long}}$ for best performance. For $h_c$ = 0.5 W/(m²K), the steady-state temperature is shown in Fig. 3(a) for different combinations of $\lambda_{\text{short}}$ and $\lambda_{\text{long}}$. Among the various potential designs, the optimal design is $\lambda_{\text{short}}$ = 8.30 μm and $\lambda_{\text{long}}$ = 12.38 μm, with a corresponding steady-state temperature of 274.40 K. Whereas the ideal, multi-band emitters depicted in Fig. 2 have non-negative net radiative emission at all wavelengths, single-band emitters have net radiative absorption at some wavelength regions, as shown in Fig. 3(b) with a red color. Nonetheless, an optimally designed single-band emitter can exhibit a considerably lower steady-state temperature ($T_{\text{ideal,SB}}$) than $T_{8-13}$ for highly insulated systems (Fig. 3(c)). In particular, at the perfect insulation limit, $T_{\text{ideal,SB}}$ and $T_{8-13}$ converge to 243.64 K and 268.31 K, respectively, exhibiting a difference of 24.67 K. Even in a more realistic case of $h_c$ = 0.13 W/(m²K), $T_{\text{ideal,SB}}$ (265.60 K) is 5 K lower than $T_{8-13}$ and the corresponding emission band is from 9.98 μm to 12.26 μm. Figure 3(d) illustrates the optimal emission band for a wide range of $h_c$ values. At high $h_c$ values, the ideal emission band for a single-band radiative cooler is similar to previous designs of 8 to 13 μm. However, at low $h_c$ values, the optimal emission band narrows down considerably. In particular, it can be seen from Figs. 3(c–d) that it is better to abandon the wavelength range from 8 to 10 μm if the target steady-state temperature is lower than the freezing point of water. Of course, if dual or multi-band designs are permissible, a part of this wavelength range can be used for radiation to decrease the steady-state temperature further or to increase the net radiative power density.

In conclusion, we presented a systematic method to calculate the ultimate lower bound of a radiatively cooled object's steady-state temperature as well as the ultimate upper bound of net radiative power density at a given cooler temperature under general environmental conditions with an arbitrary effective non-radiative heat transfer coefficient. We also derived the ideal spectral emissivities that can reach such bounds. Unlike the often-adopted contiguous emission window of 8 to 13 μm used in previous radiative coolers, the ideal radiative cooler exhibits unity emissivity over disjoint sets of wavelengths. We also investigated the ideal emission band for a single-band emitter and found that the optimal band narrows down considerably at lower temperatures. The proposed scheme may serve as a basic guideline for designing the emissive properties of extreme radiative coolers as well as for estimating the amount of thermal insulation required for them.

**Acknowledgements**

This research was supported by the National Research Foundation of Korea(NRF) grant funded by the Korea government(MSIT) (No. 2018M3D1A1058998 and No. 2017R1A2B2005702).

**References**

[1] C. Elachi and J. Van Zyl, *Introduction to the Physics and Techniques of Remote Sensing*, 2nd ed. (John Wiley & Sons, 2006).
[2] E. Rephaeli, A. Raman, and S. Fan, Nano Lett. **13**, 1457 (2013).
[3] A.P. Raman, M.A. Anoma, L. Zhu, E. Rephaeli, and S. Fan, Nature **515**, 540 (2014).
[4] M.M. Hossain, B. Jia, and M. Gu, Adv. Opt. Mater. **3**, 1047 (2015).
[5] Z. Chen, L. Zhu, A. Raman, and S. Fan, Nat. Commun. **7**, 1 (2016).
[6] Z. Huang and X. Ruan, Int. J. Heat Mass Transf. **104**, 890 (2017).
[7] J. long Kou, Z. Jurado, Z. Chen, S. Fan, and A.J. Minnich, ACS Photonics **4**, 626 (2017).
[8] C. Zou, G. Ren, M.M. Hossain, S. Nirantar, W. Withayachumnankul, T. Ahmed, M. Bhaskaran, S. Sriram, M. Gu, and C. Fumeaux, Adv. Opt. Mater. **5**, 1 (2017).
[9] E.A. Goldstein, A.P. Raman, and S. Fan, Nat. Energy **2**, 1 (2017).
[10] Y. Zhai, Y. Ma, S.N. David, D. Zhao, R. Lou, G. Tan, R. Yang, and X. Yin, Science **355**, 1062 (2017).
[11] S. Atiganyanun, J.B. Plumley, S.J. Han, K. Hsu, J. Cytrynbaum, T.L. Peng, S.M. Han, and S.E. Han, ACS Photonics **5**, 1181 (2018).
[12] G.J. Lee, Y.J. Kim, H.M. Kim, Y.J. Yoo, and Y.M. Song, Adv. Opt. Mater. **6**, 1 (2018).
[13] B. Bhatia, A. Leroy, Y. Shen, L. Zhao, M. Gianello, D. Li, T. Gu, J. Hu, M. Soljačić, and E.N. Wang, Nat. Commun. **9**, 1 (2018).
[14] D. Wu, C. Liu, Z. Xu, Y. Liu, Z. Yu, L. Yu, L. Chen, R. Li, R. Ma, and H. Ye, Mater. Des. **139**, 104 (2018).
[15] H. Bao, C. Yan, B. Wang, X. Fang, C.Y. Zhao, and X. Ruan, Sol. Energy Mater. Sol. Cells **168**, 78 (2017).
[16] Y. Iijima and M. Shinoda, J. Appl. Meteorol. **41**, 734 (2002).
[17] A.W. Harrison, Sol. Energy **26**, 243 (1981).
[18] X. Berger, J. Bathiebo, F. Kieno, and C.N. Awanou, Renew. Energy **2**, 139 (1992).
[19] J.A. Duffie, W.A. Beckman, and J. McGowan, *Solar Engineering of Thermal Processes*, 4th ed. (John Wiley & Sons, 2013).
[20] C.G. Granqvist and A. Hjortsberg, J. Appl. Phys. **52**, 4205 (1981).
[21] A. Berk, P. Conforti, R. Kennett, T. Perkins, F. Hawes, and J. van den Bosch, Proc. SPIE **9088**, 90880H-1 (2014).
[22] M.M. Hossain and M. Gu, Adv. Sci. **3**, 1 (2016).
[23] D. Zhao, A. Aili, Y. Zhai, S. Xu, G. Tan, X. Yin, and R. Yang, Appl. Phys. Rev. **6**, 021306 (2019).
[24] L. Zhu, A. Raman, K.X. Wang, M.A. Anoma, and S. Fan, Optica **1**, 32 (2014).
[25] L. Zhu, A. Raman, and S. Fan, Appl. Phys. Lett. **103**, 223902 (2013).

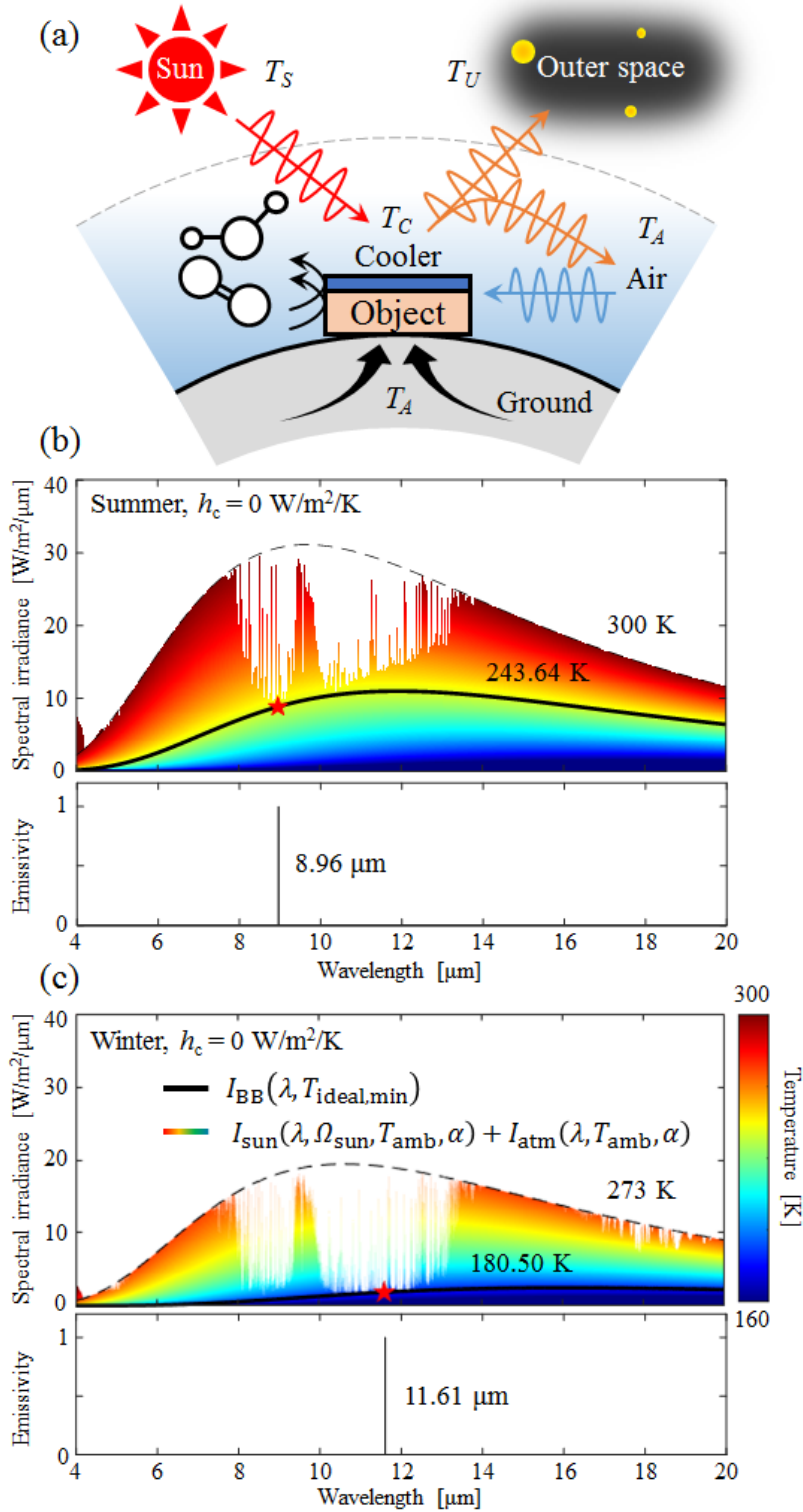

Figure 1. (a) Schematic of a radiative cooler and its surroundings. (b–c) Spectral irradiance of a blackbody at the steady state temperature of the ideal radiative cooler (black solid line) and that of the sun and the atmosphere combined (border line of the colored area) in (b) summer and (c) winter for $h_c = 0$ W/(m$^2$K) at Daejeon city. Each color indicates the spectral irradiance of a blackbody at different temperatures (black dashed lines are for ambient temperature). The bottom panels show the ideal spectral emissivities.

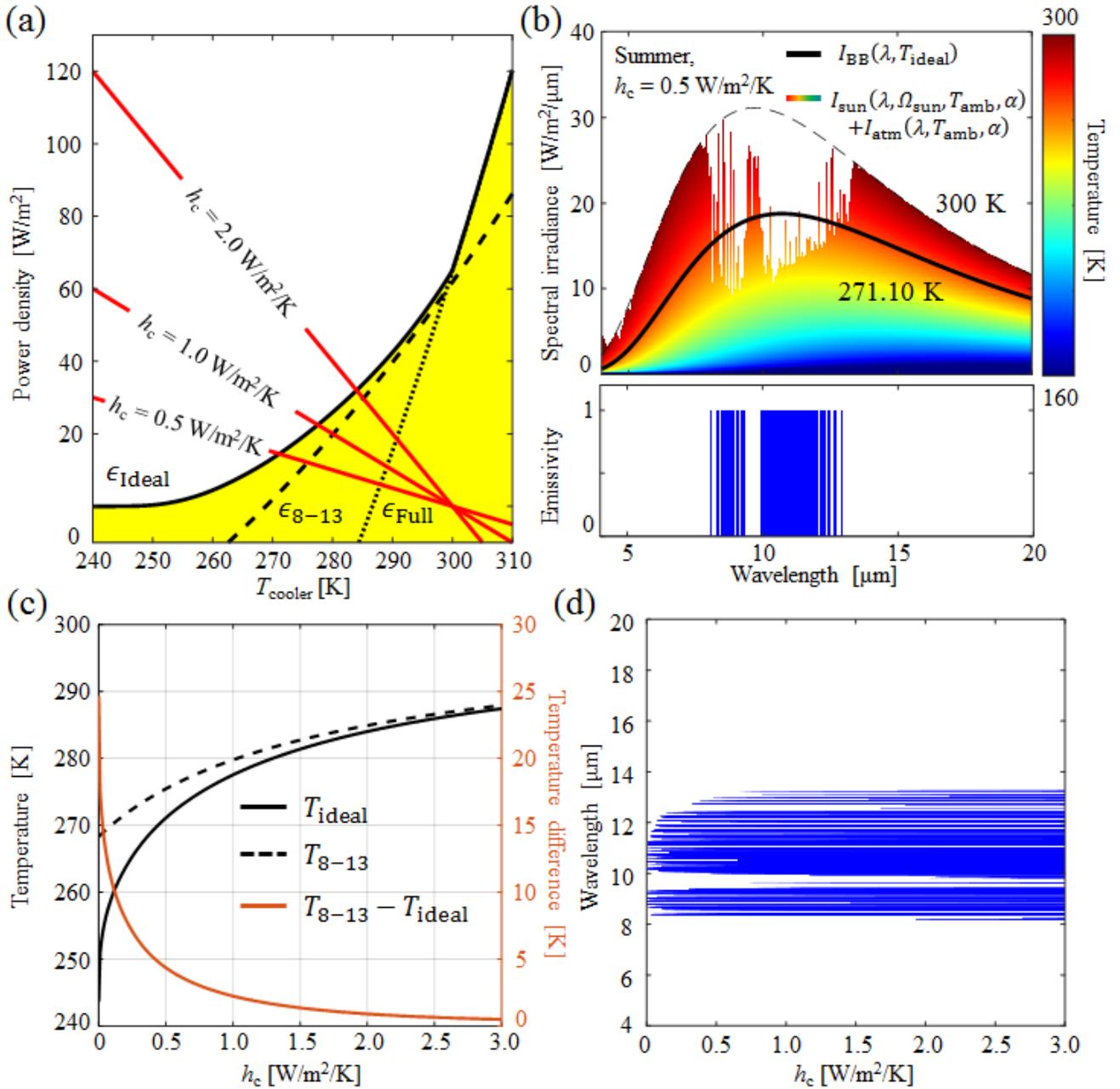

Figure 2. (a) Radiative (black) and non-radiative (red) power densities as a function of cooler temperature. (b) Spectral irradiance and spectral emissivity of the ideal radiative cooler for $h_c = 0.5$ W/(m²K). (c) The steady-state temperature as a function of $h_c$. In (a, c), data for the ideal radiative cooler is plotted with black solid line, the 8–13 μm emitter with black dashed lines, and the broadband emitter with black dotted lines. (d) Ideal spectral emissivity as a function of $h_c$.

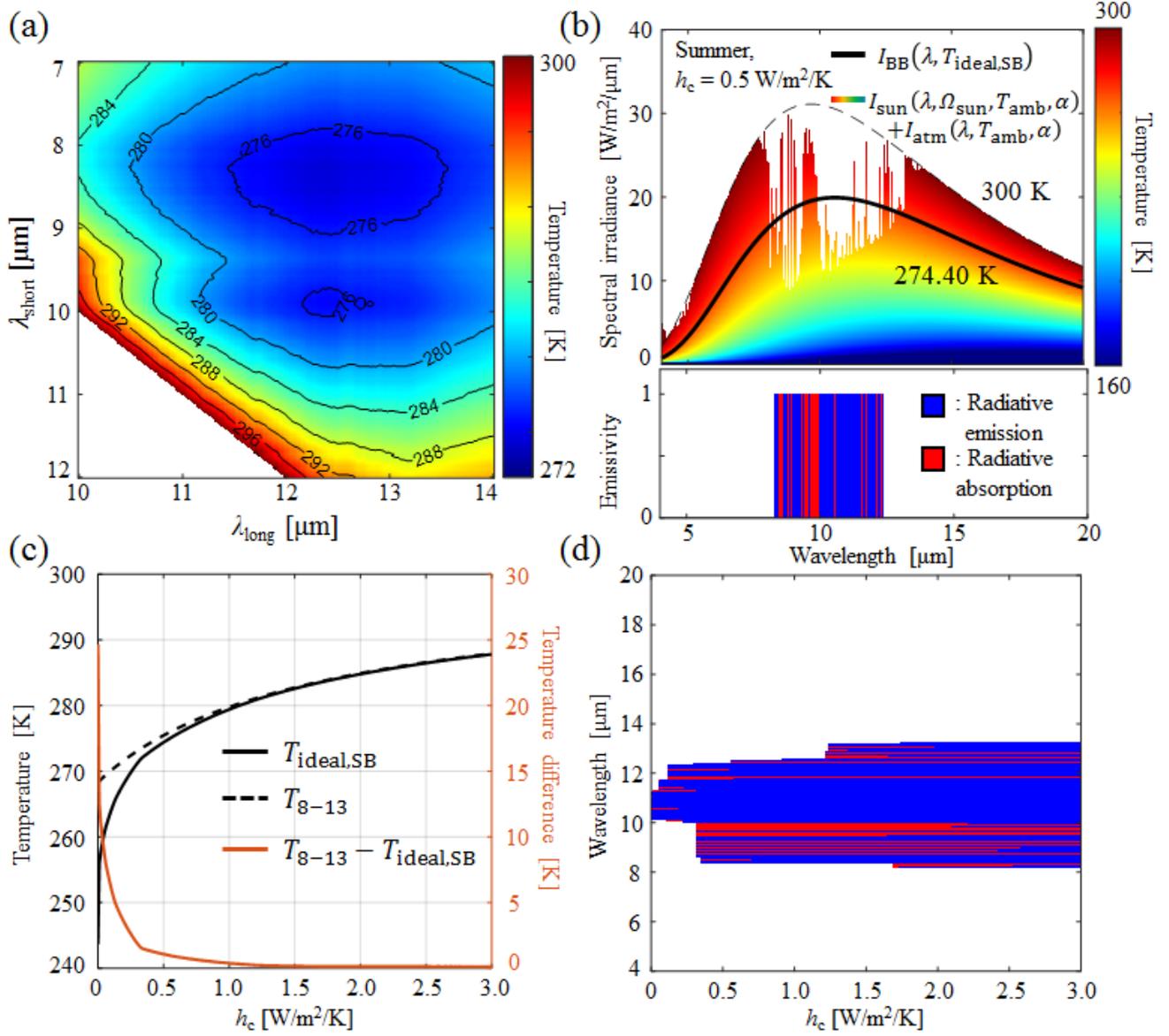

Figure 3. (a) The steady-state temperature of a single-band emitter for various $\lambda_{\text{short}}$ and $\lambda_{\text{long}}$ combinations. (b) Spectral irradiances (top panel) and the spectral emissivity of an ideal single-band emitter for $h_c = 0.5$ W/(m²K) (bottom panel). The blue and red areas in the bottom panel indicate the spectral regions of emission and absorption via radiation, respectively. (c) The steady-state temperature of the ideal single-band emitter (black solid line) and the 8–13 μm emitter (black dashed line), and the difference between them (red line), as a function of $h_c$. (d) Ideal single-band spectral emissivity as a function of $h_c$.

# Ideal spectral emissivity design for extreme radiative cooling - Supplementary material

Suwan Jeon[1] and Jonghwa Shin[1,a)]

[1]Department of Materials Science and Engineering, Korea Advanced Institute of Science and Technology, Daejeon 34141, Republic of Korea

## A. Spherical shell model

As electromagnetic wave transmitting through the atmosphere, it is scattered and absorbed by air molecules. The level of attenuation depends on atmospheric thickness along the line-of-sight, which can be represented by the angle of sight. In order to evaluate cooling performance of radiative coolers, which normally emit thermal radiation over all directions, angular properties of atmospheric transmittance must be precisely described. From Beer-Lambert law, the transmittance through lossy medium, i.e., the atmosphere, at the zenith angle $\theta$ can be expressed as $T(\theta) = 10^{-A(\theta)}$. The angle-dependent attenuation coefficient $A(\theta)$ can be approximated as $A(\theta) = A_0 \cdot h(\theta)/h_0$, where $A_0$ is the attenuation constant, $h(\theta)$ is the angle-dependent thickness of the atmosphere, and $h_0$ is the atmospheric thickness at the zenith direction. Then, the angle-dependent transmittance through the atmosphere can be arranged as

$$T(\theta) = 10^{-A(\theta)} = (10^{-A_0})^{h(\theta)/h_0} = (T_0)^{AM(\theta)} \tag{S1}$$

where $T_0$ is the atmospheric transmittance at the zenith direction (denoted as $t(\lambda, T_{amb}, \alpha)$ in the manuscript to express spectral and environmental effects) and $AM(\theta) = h(\theta)/h_0$ is the air mass coefficient at the zenith angle $\theta$. In flat earth model, mostly used in previous studies, the air mass coefficient is $AM(\theta) = 1/\cos\theta$. Although it might be a simple model expressed only with zenith angle, the air mass coefficient diverges at large zenith angle, close to 90°. On the other hand, spherical shell model can represent atmospheric thickness even at large zenith angle by assuming that the atmosphere surrounds the earth just like a spherical shell (Fig. S1(a)). In this paper, we applied geometric parameters as $h_0 = 99$ km and $R = 6400$ km. Figure S1(b) illustrates that spherical shell model is reliable over 70° contrary to flat earth model.

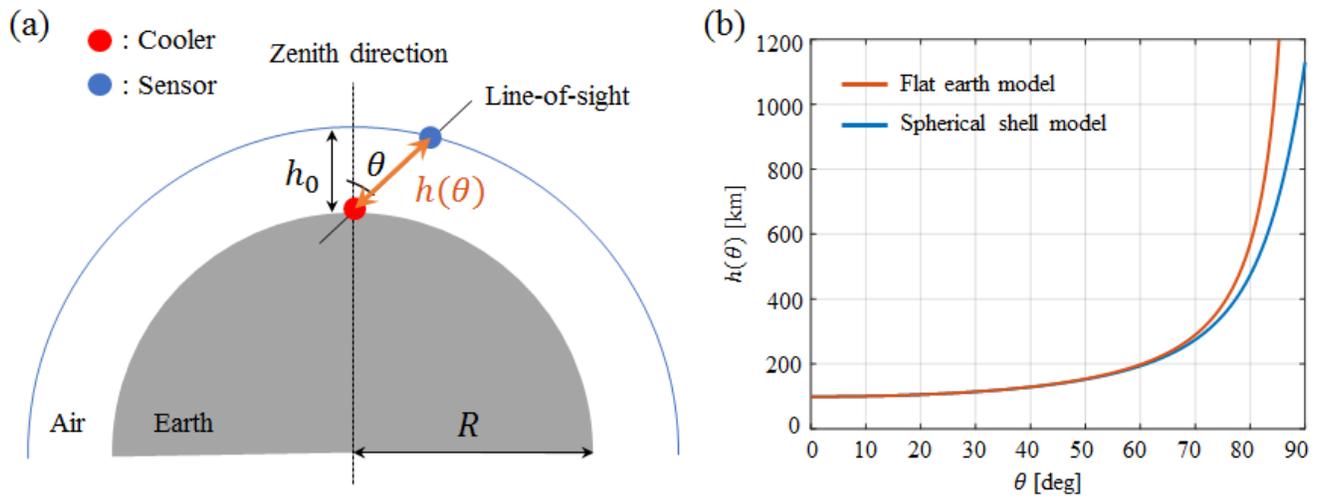

Figure S1. (a) Schematic of spherical shell model and (b) atmospheric thickness as a function of the zenith angle for flat earth model and spherical shell model.